\documentclass[onecolumn,nofootinbib,preprintnumbers,showkeys]{revtex4-2}
\usepackage{amssymb}
\usepackage{amssymb}
\usepackage{mathrsfs}
\usepackage{float}
\usepackage{bm}
\usepackage{amsmath}
\usepackage{mathdots}
\usepackage{graphicx}
\usepackage{subfigure}
\usepackage{array}
\usepackage{lipsum}
\usepackage{color}
\usepackage{hyperref}
\usepackage{ulem}
\begin{document}
	
	\title{Tangential Casimir force in the misaligned system: Magnetic media, real conductors, and a torque}
	
	\newcommand*{\SDUST}{Department of Physics and Institute for Theoretical Physics,\\Shandong University of Science and Technology, Qingdao, Shandong 266590, China}\affiliation{\SDUST}

	\author{Zhentao Zhang}\email{zhangzt@sdust.edu.cn}\affiliation{\SDUST}

	\begin{abstract}	
		  Uncharged parallel plates in the misaligned system can experience a tangential Casimir force between them. We consider the role of magnetic response in this effect by extending the tangential force to magnetic media, and the extension is realized by working out the total zero-point energy of multilayered magnetodielectrics. Then we investigate the tangential force for real conductors by taking into account the temperature dependence of their dielectric constants, and obtain needed results for experimental investigations that are expected to be conducted at room temperature. Thereafter, we discuss a Casimir torque between parallel plates made of isotropic media, which offers a simple way to realize torques for uncharged surfaces.
	\end{abstract}


\maketitle

\section{Introduction}

In 1948 Casimir predicted that two parallel, perfectly conducting plates at small separations would experience an attractive force between them, due to the modification of zero-point energy of the radiation field in the configuration~\cite{Casimir}. Thereafter, the interactions between uncharged dielectric surfaces were considered by Lifshitz and the result of Casimir can be found in the perfectly conducting limit in Lifshitz's theory~\cite{Lifshitz}. The Casimir force was demonstrated in a conclusive experiment by Lamoreaux in 1997~\cite{Lamoreaux}, which has stimulated both theoretical and experimental studies of the Casimir effect. Remarkable progress since then has been made in this field, see,~e.g., Refs.~\cite{Bordag,Lamoreaux2005,Klimchitskaya,Woods} for reviews. 

In modern experiments of the Casimir force, we usually measure the force between a sphere and a plate~\cite{Lamoreaux,Mohideen}, where the sphere-plate configuration can be related to a parallel configuration by the Derjaguin approximation~\cite{Derjaguin} but could avoid difficulties in assuring the parallelism of two plates. Nevertheless, a measurement of the force in the plate-plate configuration was successfully carried out~\cite{Bressi}. Beyond the parallel plates, researchers observed the lateral Casimir force caused by the corrugated surfaces with uniaxial corrugations of equal period~\cite{Golestanian1,Golestanian2,Chen1,Chen2}. To our knowledge, however, all the experiments performed so far for the (sphere-related) parallel conducting plates, the most elementary Casimir objects, were dedicated to the normal Casimir force which is perpendicular to the surfaces. Recently, we discussed that there may be a tangential Casimir force between parallel conducting plates in the misaligned system~\cite{Zhang}. The tangential force has unusual properties. For example, in most of experimental parameter space of a representative configuration for \textit{macroscopic}~\cite{Zhang} plates, the strength of this force is effectively independent of the overlapping area of the plates.\footnote{See the configuration in figure 2 and its variants therein. The occurrence of this property in the configuration is associated with the validity of the ideal boundary approximation that considers the Casimir effect for the overlapping areas of plates as for infinite planes, i.e., ignoring the edge effect. We here refer the reader to Ref.~\cite{Zhang} for a thorough discussion on the reliability of applying this approximation to the misaligned systems considered by us. This approximation will also be employed in the present study, which would be sufficient for our purposes.} The force was then generalized to dielectrics and finite temperature~\cite{Zhang}.

In view of nontrivial properties and implications of the tangential force, in this paper we deal with some major aspects of the effect. To consider the role of magnetic properties in this effect, in section II we shall extend the tangential force to magnetic media. Generally, below that the near-infrared and visible electromagnetic spectra which are the important frequency regions to Casimir physics, permeabilities of magnetic media would already be unity~\cite{Landau}. Thus, aside from purely theoretical explorations~\cite{Richmond,Langbein} which certainly are important, it might be appropriate to say that for a long time there were no strong realistic needs for investigating the Casimir effect in magnetic media. But the recent interest in the normal Casimir force affected by magnetic properties has grown, since it was gradually realized that the magnetic response can indeed play a valuable role in the Casimir effect~\cite{Kenneth,Iannuzzi,Tomas,Buhmann,Rosa,Banishev,Banishev2,Bimonte,Klimchitskaya1}. Hence, the present extension for magnetic media is justified. From a technical point of view, in this extension a major obstacle that we have no established concept for the total zero-point energy of the multilayered system has been removed~\cite{Zhang}.  

In section III we shall focus on the theoretical predictions for real metals, which will be essential for experimental verification of the force. In the previous study, the force at zero temperature for gold plates was investigated with its optical data~\cite{Palik}. However, a nontrivial factor that the temperature dependence of the dielectric constant was not taken into account, see, for example, Ref.~\cite{Winsemius}. Certainly, to obtain reliable results, the finite-temperature formulation should be employed if we use the optical data measured at room temperature. Besides this self-consistency issue, there is a more important reason for presenting a detailed study of the finite-temperature predictions on real conductors: Casimir experiments are usually performed at room temperature. Thus, in this section we shall investigate the tangential force at finite temperature for real metals, gold, aluminum, and ferromagnetic metal nickel, by considering their dielectric properties at room temperature. 

For parallel plates, there are three types of motion between them: normal, tangential, and rotational motions, influenced by normal forces, tangential forces, and torques, respectively. Thus, from the perspective of mechanical degrees of freedom, it is of interest to realize torques between uncharged parallel plates of isotropic media. And we shall discuss a torque of this type in our section IV. It should be mentioned that there is a known Casimir torque for parallel plates of \textit{anisotropic} media~\cite{Kats,Parsegian,Barash}. Recently, this torque was confirmed experimentally~\cite{Somers}. Besides anisotropic media and flat surfaces, Casimir torques were also considered for infinitely long, cylindrical thin rods separated by distances much greater than their diameters~\cite{Parsegian2,Hopkins} and for periodically corrugated surfaces~\cite{Rodrigues,Rodrigues2}, see Ref.~\cite{Munday} for a recent review and further references.

\section{The tangential force in magnetic media}

\begin{figure}[H]
	\centering
	\includegraphics[width=5.4cm]{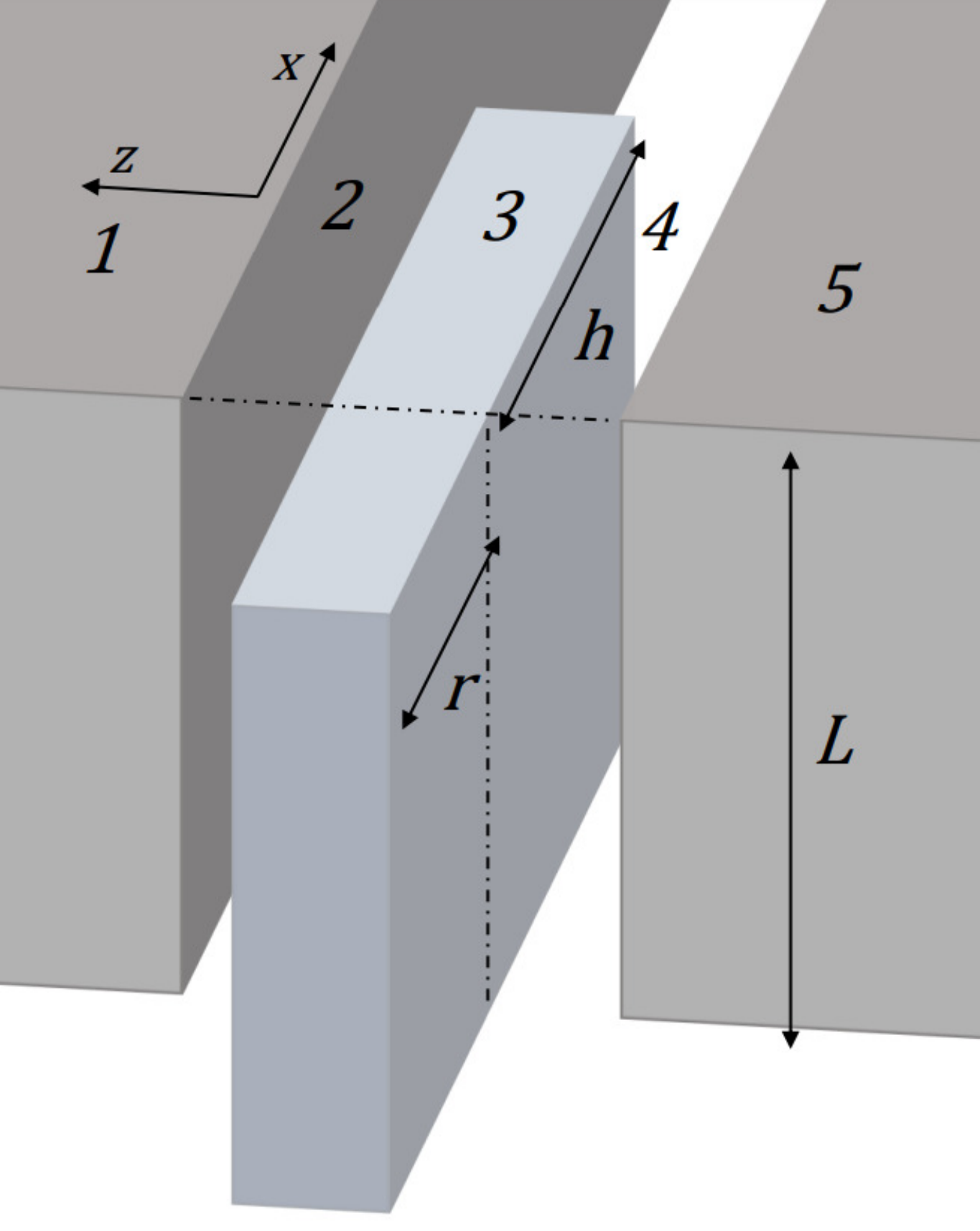}
	\caption{Two slabs and a smaller plate separated by other media. It is a multilayered system in which isotropic media occupy the five regions in the overlapping area. Thicknesses of the three planar layers are $d_{2,3,4}$, and lengths of the inside and the outside parts of the middle plate are denoted by $h$ and $r$. We shall ignore the edge contribution in this system by assuming that their widths $L>10^{-4}$ m, i.e., macroscopic plates, and $\text{min}\{d_2,d_4\}\lesssim 10^{-6}~\text{m}\ll \text{min}\{h,r\}$, which would be an ordinary experimental setting in Casimir physics~\cite{Zhang}.} 
	\label{magsys}
\end{figure}

Consider the misaligned system of magnetodielectrics in Fig.~\ref{magsys}, where the widths of the parallel rectangular plates, along the $y$-axis in the Cartesian coordinate system, are the same. 

To investigate the tangential force experienced by the middle plate, one needs to find the total zero-point energy of the multilayered magnetic media. A Casimir energy between two planar magnetodielectric multilayers was discussed in Ref.~\cite{Langbein}. However, as noted in the case of dielectrics~\cite{Zhang}, we are interested in the total energy of a $n$-layer (say) system $E_0$ which is one and only for that whole system, and it could give the normal forces between layers $i-1$ and $i+1$ simply by
\begin{equation}
	F^N_{i-1i+1}=-\frac{\partial E_0(d_2,d_3,...,d_{n-1})}{\partial d_i},
\end{equation}
where $i=2,3,...,n-1$. 

To find the total energy of a five-layer magnetodielectric system as shown in Fig.~\ref{magsys}, we could employ the surface mode method \cite{van Kampen} developed in Ref.~\cite{Milonni}. Assume that $\epsilon_i(\omega)$, $\mu_i(\omega)$ $(i=1,2,3,4,5)$ are, respectively, the frequency-dependent dielectric constants and permeabilities of the media. By the surface mode approach~\cite{Milonni} with the analytical properties of the permeabilities~\cite{Richmond,Langbein,Landau}, due to the observation in Ref.~\cite{Zhang}, the total zero-point energy of the magnetodielectric system may be found to be
\begin{align}
	E_0(d_2,d_3,d_4)=&\frac{\hbar L^2}{4\pi^2}\int_{0}^{\infty}dk_\Arrowvert k_\Arrowvert\int_{0}^{\infty}d\xi\sum_{\lambda=\alpha,\beta}\ln G^{\lambda}(i\xi;d_2,d_3,d_4),
	\label{total-zero-point-energy1}
\end{align}
where the overlapping area of the layers is assumed to be $L\times L$, the in-plane wavenumber $k_{\parallel}=(k^2_x+k^2_y)^{\frac{1}{2}}$, the frequency $\omega=i\xi$, and  
\begin{align}
	G^{\lambda}(\omega)\equiv1-&\sum^{4}_{i=2}r^{\lambda}_{i-}r^{\lambda}_{i+}e^{-2K_id_i} -\sum^{3}_{i=2}r^{\lambda}_{i-}r^{\lambda}_{(i+1)+}e^{-2K_id_i-2K_{i+1}d_{i+1}} \nonumber \\ +&r^{\lambda}_{2-}r^{\lambda}_{2+}r^{\lambda}_{4-}r^{\lambda}_{4+}e^{-2K_2d_2-2K_{4}d_{4}} -r^{\lambda}_{2-}r^{\lambda}_{4+}e^{-2K_2d_2-2K_{3}d_{3}-2K_{4}d_{4}},~~\lambda=\alpha,\beta,
	\label{surface-modes-condition}
\end{align}
where 
\begin{align}
	r^{\alpha}_{i+}=\frac{\mu_{i+1}K_{i}-\mu_{i}K_{i+1}}{\mu_{i+1}K_{i}+\mu_{i}K_{i+1}}=-r^{\alpha}_{(i+1)-},~~r^{\beta}_{i+}=
	\frac{\epsilon_{i+1}K_{i}-\epsilon_{i}K_{i+1}}{\epsilon_{i+1}K_{i}+\epsilon_{i}K_{i+1}}=-r^{\beta}_{(i+1)-},
	\label{de}
\end{align}
and
\begin{align}
 K_i=\sqrt{k_\Arrowvert^2-\epsilon_{i}(\omega)\mu_{i}(\omega)\frac{\omega^2}{c^2}}.
 \label{de2}
\end{align}
Extending the total energy to any-number magnetodielectric layers would be the same as the one detailed for dielectrics~\cite{Zhang}.

From the zero-temperature formula given above, the energy at finite temperature can be found by adding a thermal contribution term $\coth[i\hbar\xi/(2k_BT)]$~$(T\neq0)$~\cite{Milonni}, where $k_B$ is the Boltzmann constant and $T$ is the temperature. And we may obtain
 \begin{align}
	E_0(d_2,d_3,d_4;T)=&\frac{k_B T L^2}{2\pi}\sum_{n=0}^{\infty}{'}\int_{0}^{\infty}dk_\Arrowvert k_\Arrowvert\sum_{\lambda=\alpha,\beta}\ln G^{\lambda}(\xi_n;d_2,d_3,d_4),
	\label{total-zero-point-energy2}
\end{align}
where $\Sigma'$ means that there is a weight factor $1/2$ for the $n=0$ term, and
 \begin{align}
\xi_n=\frac{2n \pi k_B T}{\hbar}.
	\label{frequency}
\end{align}
Notice that we here have simply written $G^{\lambda}(i\xi_n)$ as $G^{\lambda}(\xi_n)$.

Therefore, according to Ref.~\cite{Zhang}, there might be a tangential force along the $x$-axis experienced by the middle plate in the misaligned magnetodielectric system, and the force, labeled with subscript $T$, is
 \begin{align}
	F_{T}(d_2,d_3,d_4;T)=&-\frac{k_B T L}{2\pi}\sum_{n=0}^{\infty}{'}\int_{0}^{\infty}dk_\Arrowvert k_\Arrowvert \nonumber\\
	&\times\sum_{\lambda=\alpha,\beta}\left[\ln G^{\lambda}(\xi_n;d_2,d_3,d_4)-\ln G^{\lambda}(\xi_n;d_3)-\ln G^{\lambda}(\xi_n;d_2+d_3+d_4)\right],
	\label{TFatT}
\end{align}
where
 \begin{align}
    G^{\lambda}(\xi_n;d_3)&=G^{\lambda}(\xi_n;d_2,d_3,d_4)|_{d_{2,4}\rightarrow\infty}, \nonumber\\
 	 G^{\lambda}(\xi_n;d_2+d_3+d_4)&=G^{\lambda}(\xi_n;d_2,d_3,d_4)|_{\epsilon_{2}=\epsilon_{3}=\epsilon_{4},\mu_{2}=\mu_{3}=\mu_{4}}. \nonumber
\end{align}	
In general, one requires $\epsilon_{2}=\epsilon_{4}$ and $\mu_2=\mu_4$ for considering the tangential force. 

We see that magnetic responses of media come into play in the total zero-point energy and then the tangential force \textit{via} the modifications of $r^{\alpha}_{i\pm}$ and $K_i$ in Eqs.~(\ref{de}) and (\ref{de2}), compared with the original formulation of non-magnetic media in Ref.~\cite{Zhang}. It reflects in this circumstance a property of the Casimir energies in magnetic media~\cite{Richmond}.

\section{Tangential forces for real conducting plates}

\begin{figure}[H]
	\centering
	\includegraphics[width=4cm]{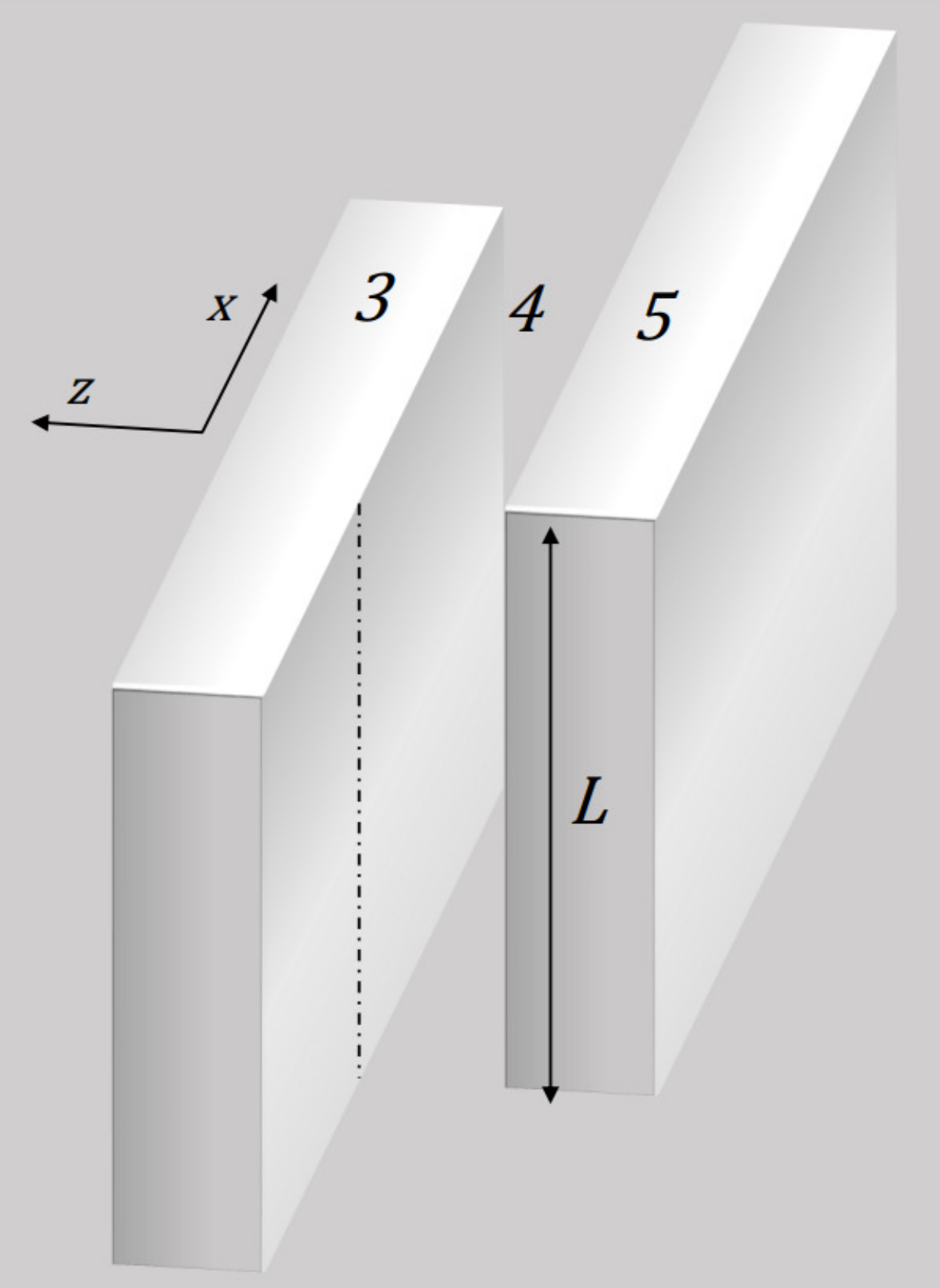}
	\caption{Two parallel but misaligned metal slabs separated by the medium in region 4. Note that for measuring the tangential force, typical metal plates can be well approximated as infinitely thick layers if their thicknesses are larger than a few hundred nanometers~\cite{Zhang}.} 
	\label{magsys1}
\end{figure}

We now discuss the tangential force for real metals with their optical data. We shall study the system of two identical plates, $d_{2,3}\rightarrow\infty$ effectively in Fig.~\ref{magsys}, for gold (Au), aluminum (Al), and nickel (Ni), see Fig.~\ref{magsys1}. Then, the tangential force in Eq.~(\ref{TFatT}) is reduced to
 \begin{align}
	F_T(d_4;T)=-&\frac{k_B T L}{2\pi}\sum_{n=0}^{\infty}{'}\int_{0}^{\infty}dk_\Arrowvert k_\Arrowvert\sum_{\lambda=\alpha,\beta}\ln\left[1-r^{\lambda}_{4-}r^{\lambda}_{4+}e^{-2K_4d_4}\right],
	\label{TFatT1}
\end{align}
and in the following discussion we will take permeabilities of Au and Al to be unity.
 
To calculate the force for real conductors, we need the information on the dielectric constants on the imaginary frequency axis. The requisite values of $\epsilon(i\xi)$ can be found by a Kramers-Kronig relation~\cite{Lambrecht}, using the optical data~\cite{Palik} and an extrapolation for low frequencies below the measured range by the Drude model
\begin{equation}
	\epsilon(\omega)=1-\frac{\omega^2_p}{\omega(\omega+i\gamma)},
\end{equation}
where the plasma frequency $(\hbar\cdot)\omega_p=9.0$~eV and the relaxation frequency $\gamma=0.035~\text{eV}$ for Au.

Employing the optical data measured at room temperature, we shall assume $T=300$~K for plates separated by the vacuum ($\epsilon_{4}=\mu_{4}=1$), and then $\xi_{1}=2.468\times10^{14}$~rad/s. But it should be noted that we shall also calculate Eq.~(\ref{TFatT1}) by different consideration in which the Drude-like dielectric constant is replaced with the dielectric function of the plasma model (with $\gamma=0$ and the same $\omega_p$) for the $n=0$ term. The two treatments, the Drude and the plasma, would result in different predictions on the Casimir effect at finite temperature~\cite{Bostrom}, due to the fact that at zero frequency ($\xi_0=0$),
\begin{align}
r^{\alpha}_{4-}=r^{\alpha}_{4+}=0, ~~~r^{\beta}_{4-}=r^{\beta}_{4+}=1
\end{align}
for the Drude model whereas
\begin{align}
r^{\alpha}_{4-}=r^{\alpha}_{4+}=\frac{k_\Arrowvert-\sqrt{\smash[b]{k^2_\Arrowvert+\omega^2_p/c^2}}}{k_\Arrowvert+\sqrt{\smash[b]{k^2_\Arrowvert+\omega^2_p/c^2}}},~~~~~r^{\beta}_{4-}=r^{\beta}_{4+}=1
\end{align}
for the plasma model. However, it seems that most of experiments in this direction are in favor of the plasma treatment, though the Drude model should be a more realistic description for real metals, see Ref.~\cite{Mostepanenko} for a review on this puzzle. 

\begin{figure}[H]
	\centering
\subfigure[ The black one is the force per unit length calculated by using the Drude treatment, and the blue one (dashed) by using the plasma-like dielectric constant at zero frequency.]{\label{Au1}\includegraphics[width=8cm]{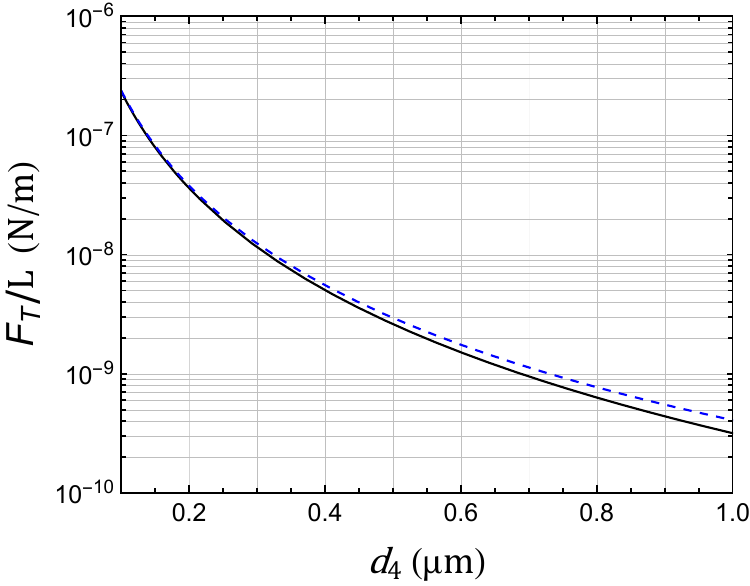}}
~~~~~~~~~\subfigure[The ratio ($R$) of the results in the left figure to the force in Ref.~\cite{Zhang}.]{\label{Au2}\includegraphics[width=7.65cm]{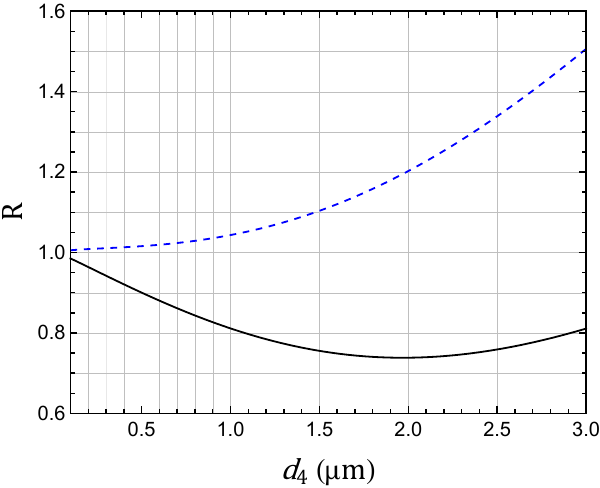}} 
\caption{The tangential force $F_T(d_4;T)$ for gold plates separated by the vacuum is calculated with the optical data in Ref.~\cite{Palik} at $T=300$ K. }
\label{Aus}
\end{figure}

Due to the rapid decay of the contribution from frequencies above the visible spectrum, we may calculate the tangential force with the frequency region up to $\xi_{n=500}$ for the separations ranging from $0.1~\mu$m to $1~\mu$m, and it is checked that the largest decrease in the results for gold plates is less than $0.0001\%$ if $n_{\text{max}}=100$ is used, see Fig.~\ref{Au1}. In Fig.~\ref{Au2}, we compare these results with the predictions obtained by the zero-temperature formula~\cite{Zhang}, and in the comparisons we also consider larger separations, which should be helpful to show the influence of thermal effect on the tangential force since the thermal effect is also dependent on the distance~\cite{Milonni}. We see that the finite-temperature Drude results significantly alter the ``zero-temperature'' results of the tangential force at both small and large separations. The differences between the finite-temperature plasma results and the previous predictions are modest at small separations, and become significant when the separation is increased.

\begin{figure}[H]
	\centering
	\subfigure[ ]{\label{Al1}\includegraphics[width=8cm]{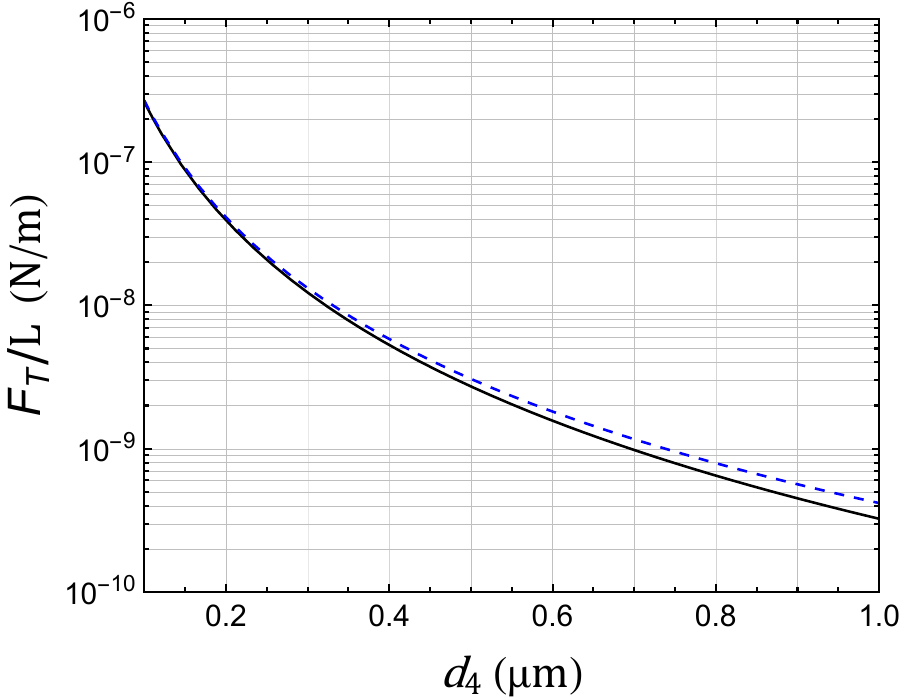}} 	~~~~~~~~\subfigure[]{\label{Al2}\includegraphics[width=7.8cm]{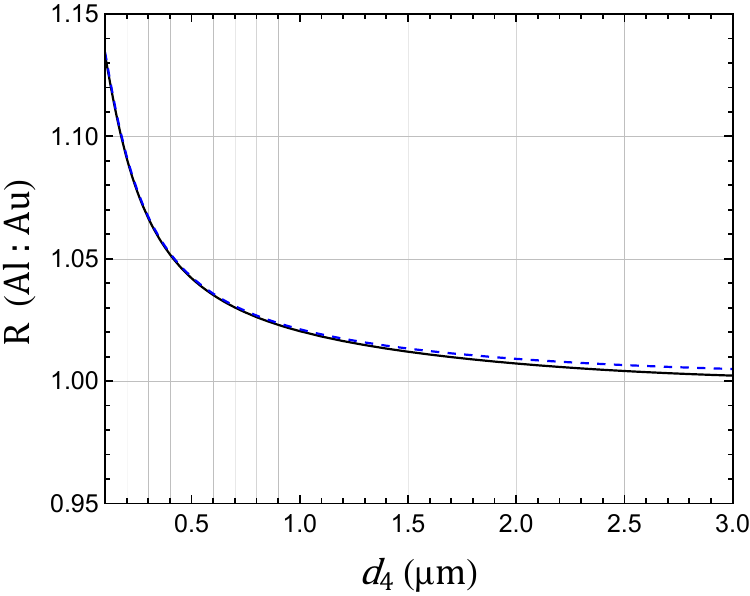}} 
	\caption{The tangential force $F_T(d_4;300~\text{K})$ per unit length calculated for aluminum plates. In Fig.~\ref{Al1}, the black one representing the results from the Drude treatment and the blue one (dashed) from the plasma treatment. To see the differences between Al-Al plates and Au-Au plates, ratios of these results to those for gold plates are shown in Fig.~\ref{Al2}. At large separations the ratios approach $1$, mainly due to the fact that the zero frequency contribution dominates in this circumstance and the parameter-dependent behaviors of the $n=0$ terms in the plasma treatment is suppressed.}
	\label{Als}
\end{figure}

We next consider the tangential force for the plates made of aluminum. To find the force, we use the optical data of Al in Ref.~\cite{Palik}, and parameters $\omega_p=12.5$~eV and $\gamma=0.063~\text{eV}$ are used to extrapolate the dielectric constant in the region below the measured frequencies. The results for aluminum plates separated by the vacuum are presented in Fig.~\ref{Als}.

\begin{figure}[H]
	\centering
	\subfigure[The ratio of the force with Au2 data to the one with Au1 data, where the black one is from the Drude treatment and the blue one (dashed) is from the plasma treatment. ]{\label{Au2vsAu1}\includegraphics[width=8cm]{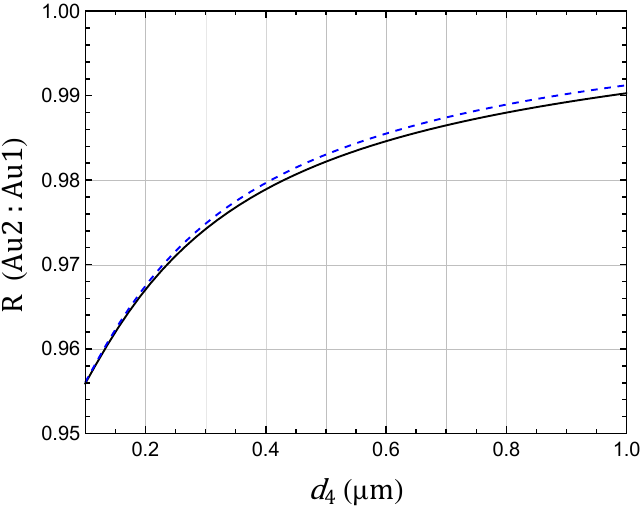}} 
	~~~~~~~~~\subfigure[The ratio of the results from Au2 data with the plasma treatment (Au2 p) to those from Au1 data with the Drude treatment~(Au1 d).]{\label{Cross}\includegraphics[width=8cm]{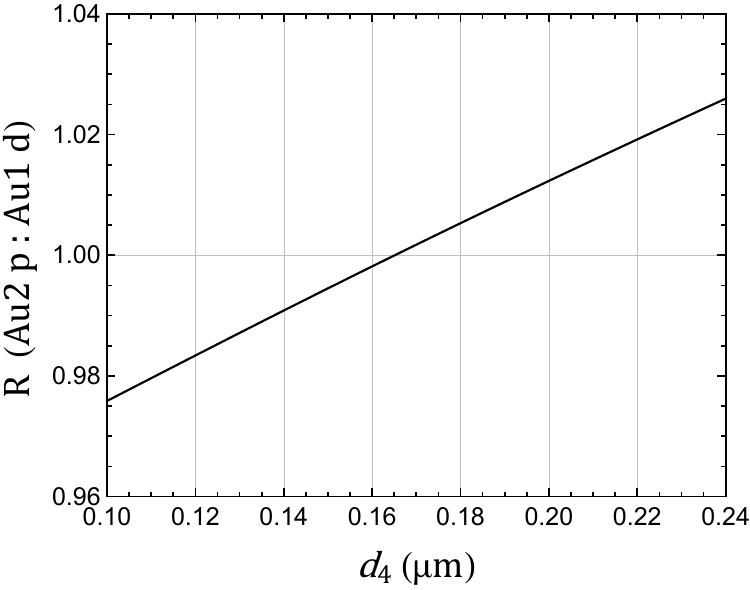}}
	\caption{A consequence of the uncertainty in optical data for the tangential force at $T=300$ K. } 
	\label{AuVs}
\end{figure}

However, we should mention that, by using optical data in any related calculation, an error source is introduced into the result, due to discrepancies in accessible optical data in the literature~\cite{Olmon}. To examine the influence of this uncertainty in optical data on the tangential force, we shall now calculate the force for gold plates with the optical data, in the range of $0.04974$ eV to $4.133$ eV, provided by Ref.~\cite{Olmon}, hereinafter referred to as the ``Au2''. In Ref.~\cite{Olmon}, three samples (evaporated gold, template-stripped gold, and single-crystal gold) representing bulk gold were used, and the differences between the optical data of them are small. The data of Au in Ref.~\cite{Palik}, hereinafter referred to as the ``Au1'', will be replaced with Au2 data of the evaporated gold below $4.2$~eV. In the calculations we shall use $\omega_p=8.45$~eV and $\gamma=0.047$~eV for extrapolating the dielectric constant at low frequencies below $0.04974$~eV. The differences between the results by using Au1 and Au2 data, respectively, are presented in Fig.~\ref{AuVs}. We see that at small separations the uncertainty in optical data could make it difficult to compare precision measurement results of the tangential force with theory. 

We now investigate the tangential force in the magnetodielectric system for real metals, by considering the magnetic responses in Eq.~(\ref{TFatT1}). Since, as shown above, the thermal effect could cause significant corrections to the tangential force in experiment, we shall also focus on the magnetic system at room temperature.
	
Consider the plates are made of Ni in Fig.~\ref{magsys1}. As explained in the introduction, for frequencies $\xi_n(n\neq0)$ at room temperature, the permeability of Ni $\mu(i\xi_n)=1$. However, due to the magnetic responses of the medium, at zero frequency one has
		\begin{align}
			r^{\alpha}_{4-}=r^{\alpha}_{4+}=\frac{\mu(0)-1}{\mu(0)+1}
		\end{align}
		for the Drude model, and 
		\begin{align}
			r^{\alpha}_{4-}=r^{\alpha}_{4+}=\frac{\mu(0)k_\Arrowvert-\sqrt{\smash[b]{k^2_\Arrowvert+\mu(0)\omega^2_p/c^2}}}{\mu(0)k_\Arrowvert+\sqrt{\smash[b]{k^2_\Arrowvert+\mu(0)\omega^2_p/c^2}}}
		\end{align}
for the plasma model. For Ni, $\mu(0)=110$ may be used in Casimir physics, see, for example, Refs.~\cite{Banishev,Banishev2,Bimonte}. To find the dielectric constant of Ni on the imaginary frequency axis, we use the optical data in Ref.~\cite{Palik}, but the data below $8$~eV are replaced with the recent, improved measurement in Ref.~\cite{Cahill}, which measures the optical constants in a broad range of $0.345$ to $7.75$ eV, and the parameters $\omega_p=4.89~\text{eV}$ and $\gamma=0.0436~\text{eV}$~\cite{Ordal} are used for extrapolating the dielectric constant at lower frequencies.

\begin{figure}[H]
	\centering
	\subfigure[The force calculated by the Drude treatment (black) and the one by the plasma treatment (blue dashed).]{\label{Nia}\includegraphics[width=8cm]{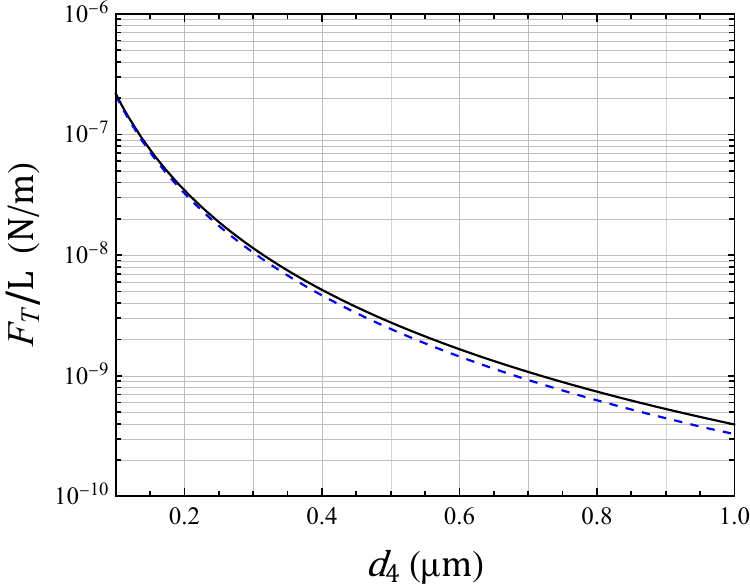}} 
	~~~~~~~~~\subfigure[{The ratio of the results in the left figure, respectively, to those from ignoring the magnetic properties of Ni [i.e., $\mu(0)=1$]}.]{\label{Nib}\includegraphics[width=7.69cm]{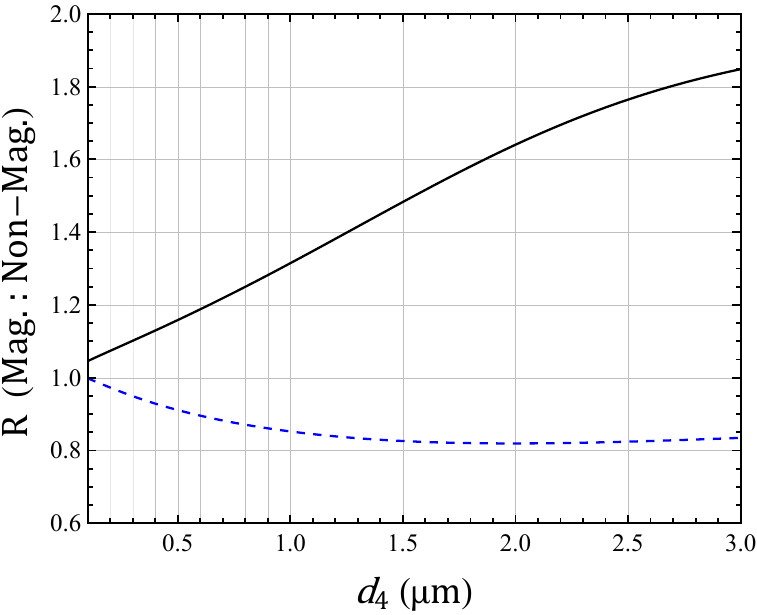}}
	\caption{The tangential force per unit length for nickel plates separated by the vacuum at $T=300$~K.} 
	\label{Ni}
\end{figure}

The numeric results of the tangential force for Ni plates are presented in Fig.~\ref{Nia}. It is seen that the force calculated by the Drude treatment is larger than the plasma treatment, due to the magnetic responses of Ni. To further manifest the influence of the magnetic properties on the force, the results are compared with the ones that ignore the magnetic responses, see Fig.~\ref{Nib}. 

\section{The Casimir torque for isotropic plates}
We now realize a Casimir torque for isotropic plates in the parallel configuration, which, to a certain extent, is a related effect of the tangential Casimir force.\footnote{This torque for perfectly conducting plates was noticed by Ref.~\cite{Zhang} [see a passing mention below Eq.~(23) therein], but the reference skipped over the torque with no information provided. This unsatisfactory situation was partly due to the fact that the fundamental role of the torque in (Casimir) mechanical systems was not fully appreciated by the author at that time.}

\begin{figure}[H]
	\centering
	\subfigure[The vertical view for the parallel plates, where $K>L$ and $H^2~>~K^2+~L^2$.]{\label{ta}\includegraphics[width=5cm]{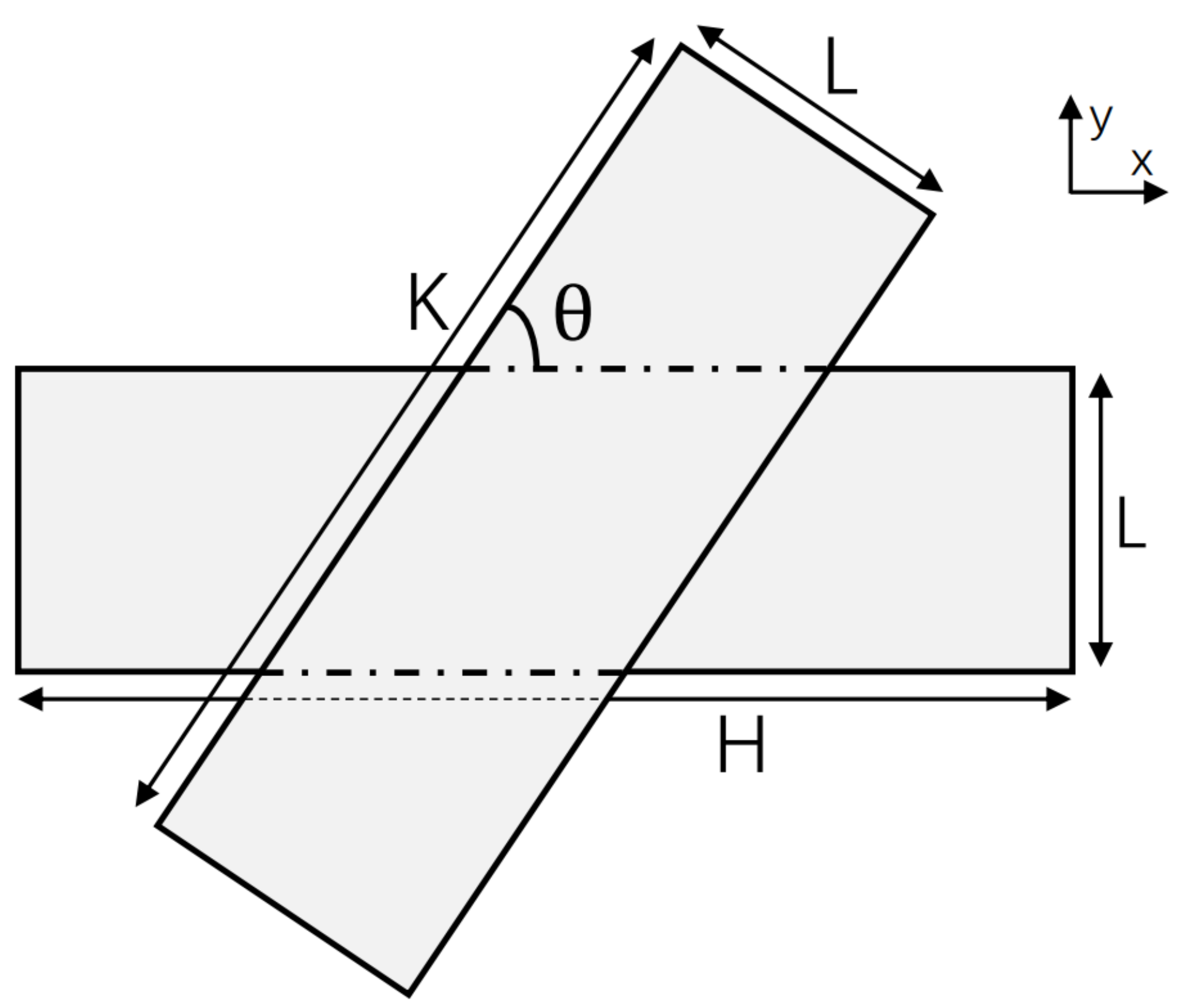}} ~~~~~~~~~~~~~~~~~~~~\subfigure[The side view (from a distance).]{\label{tb}\includegraphics[width=4.6cm]{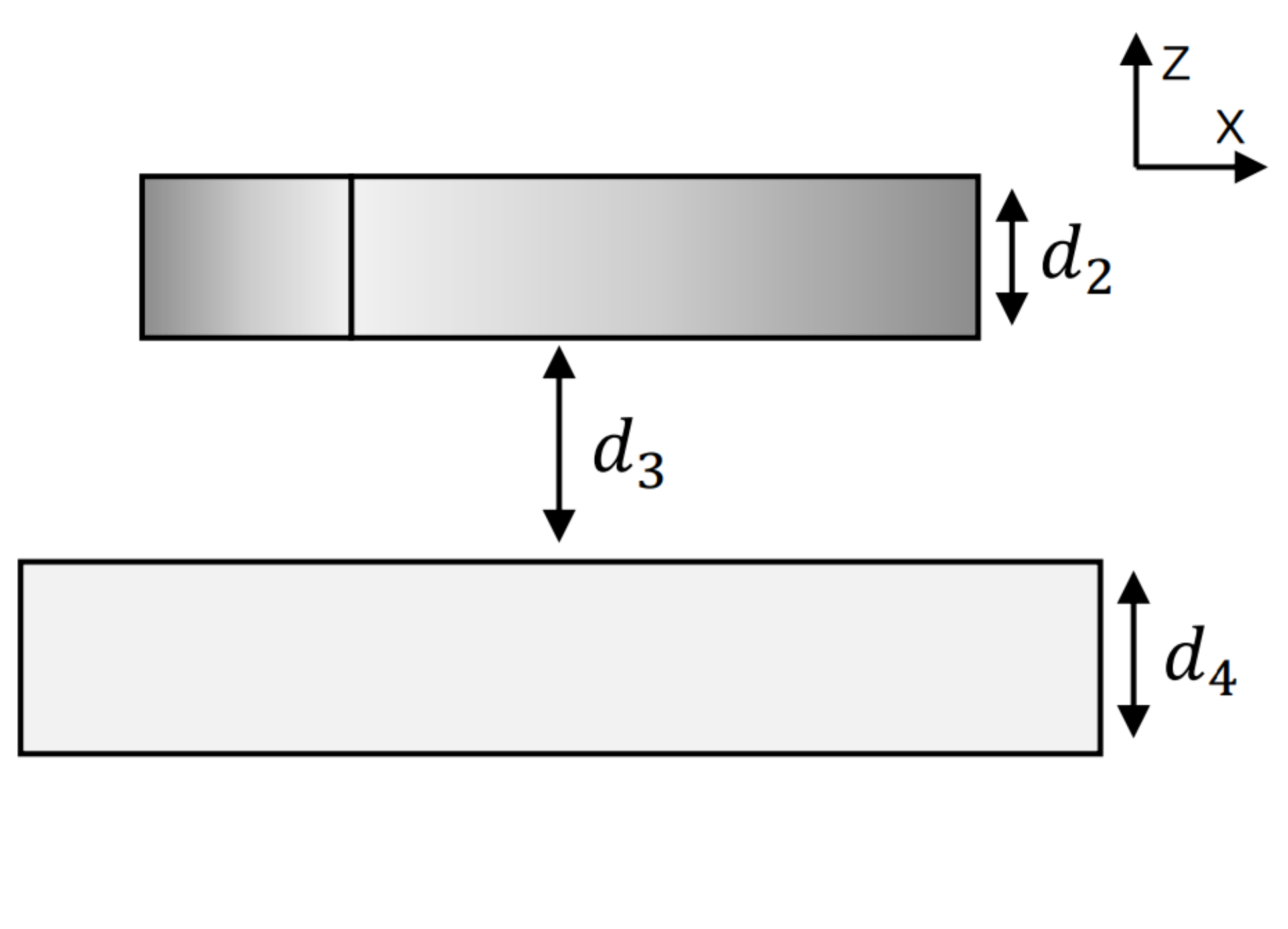}} 
	\caption{The system consists of the crossed rectangular plates of isotropic media, where the line connecting their centroids is perpendicular to the surfaces, and the widths of the plates are the same. }
	\label{T1}
\end{figure}

Consider the system in Fig.~\ref{T1}, and assume that the characteristic sizes of the plates are much larger than the distance between them, which is a typical setting for Casimir experiments.

The energy difference between the situations that the two plates are overlapped and not overlapped is
\begin{align}
	E&=E_0(d_2,d_3,d_4;T)-E_0(d_2;T)-E_0(d_4;T)~\nonumber\\
	 &=\frac{k_B T S}{2\pi}\sum_{n=0}^{\infty}{'}\int_{0}^{\infty}dk_\Arrowvert k_\Arrowvert\sum_{\lambda=\alpha,\beta}\left[\ln G^{\lambda}(\xi_n;d_2,d_3,d_4)-\ln G^{\lambda}(\xi_n;d_2)-\ln G^{\lambda}(\xi_n;d_4) \right],
\end{align}
where $E_0(d_{2(4)};T)=E_0(d_2,d_3,d_4;T)|_{d_{3,4(2)}\rightarrow\infty}$, and $S$ is the overlapping area of the plates. For the five-layer system under consideration, all the other places are occupied by the same medium ($\epsilon_1=\epsilon_3=\epsilon_5$ and $\mu_1=\mu_3=\mu_5$).

In this misaligned system the plates should experience no tangential force, i.e.,
\begin{align}
	 \textbf{F}_T=0.
\end{align}
But the Casimir energies and their difference for the plates are dependent on the angle $\theta$, i.e., there are no rotational symmetries for the energies. Thus, we may find that there is a torque experienced by the plates as
\begin{align}	
	M &=-\frac{\partial E(\theta)}{\partial \theta}\nonumber \\
	  &=-S'(\theta)\frac{k_B T}{2\pi}\sum_{n=0}^{\infty}{'}\int_{0}^{\infty}dk_\Arrowvert k_\Arrowvert\sum_{\lambda=\alpha,\beta}\left[\ln G^{\lambda}(\xi_n;d_2,d_3,d_4)-\ln G^{\lambda}(\xi_n;d_2)-\ln G^{\lambda}(\xi_n;d_4) \right],
	\label{Torque}
\end{align}
where
\begin{align}
	S'(\theta)=\frac{\partial S}{\partial \theta}=\left\{\begin{array}{c c c c}
		&\displaystyle-L^2\frac{\cos\theta}{\sin^2\theta},  &\displaystyle\theta_0<\theta\leqslant\frac{\pi}{2}  \\ [12pt]
		&\displaystyle\left(\frac{K}{2\cos\theta}-\frac{L}{\sin2\theta}+\frac{L}{2\sin\theta}\right)\left(\frac{L}{2\sin\theta}-\frac{K}{2\cos\theta}-L\cot2\theta\right),~~ &\displaystyle 0<\theta\leqslant\theta_0 \\
	\end{array}\right.%
\end{align}
in which
\begin{equation}
	\theta_0=\arcsin\frac{2KL}{K^2+L^2},
\end{equation}
with the line passing through the centroids being the rotation axis.

\begin{figure}[H]
	\centering
	\includegraphics[width=8.5cm]{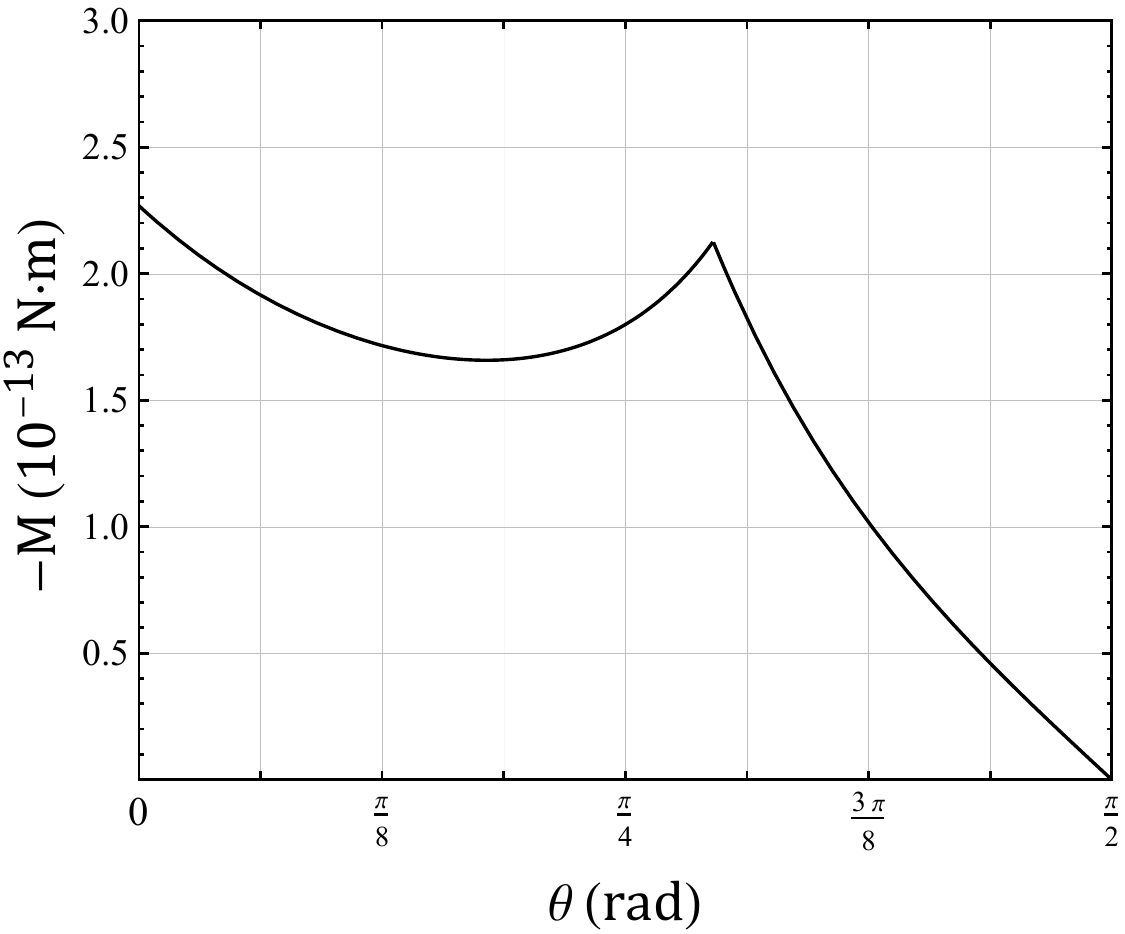}
	\caption{The torque for gold plates separated by the vacuum, where $L=1$~mm, $K=2$~mm, and $d_3=100$ nm. The non-zero value ``at'' $\theta=0$ should be understood in the sense of $\theta\rightarrow0^{+}$, since $S(\theta)$ does not have the partial derivative at $\theta=0$, where the torque vanishes.} 
	\label{TorqueAu2p}
\end{figure}

We may calculate the torque for gold plates with moderate settings at room temperature using Au2 data by the plasma treatment, see the results in Fig.~\ref{TorqueAu2p}, where we assume that the thicknesses of the plates are not very thin and then can be considered infinite in practical calculations. 

Notice that the torque in Eq.~(\ref{Torque}) is formulated with the ideal boundary assumption, i.e., ignoring the edge contributions. To examine its validity for the torque system with typical settings, we shall now estimate edge corrections to the torque. To do that, the numeric simulations of the edge effect for infinite, perfectly conducting plates~\cite{Gies,Graham} may be used as quantitative estimates for real metals, since, for our purpose, only the \textit{relative} strength of the edge contribution matters.

The perimeter of the overlapping area between the plates, $P$, also depends on $\theta$, and characteristic sizes of the area are much larger than $d_3$. Then, we may apply the misaligned edge energy at zero temperature~\cite{Graham} to the perimeter as
\begin{align}
\delta E\approx0.0009\frac{\hbar c}{d_3^2} P(\theta)
\label{edgeE}
\end{align}
when $\theta$ is not close to zero, since the scale of the dominant energy density transition region in the parallel configuration should be very small~\cite{Gies,Zhang}. This energy induces a torque $\delta M=-\partial \delta E/\partial \theta$.

Thus, by 
\begin{align}
	P'(\theta)&=\frac{\partial P}{\partial \theta}=\left\{\begin{array}{c c c c}
		&\displaystyle-4L\frac{\cos\theta}{\sin^{2}\theta},  &\displaystyle\theta_0<\theta\leqslant\frac{\pi}{2}  \\ [12pt]
		&\displaystyle\frac{K}{\cos^2\theta}\cdot\left(\sin\theta-1\right)+L\cdot\left(4\frac{\cos2\theta}{\sin^{2}2\theta}-2\frac{\cos\theta}{\sin^{2}\theta}+\frac{1}{\sin^2\theta}+\frac{\sin\theta}{\cos^{2}\theta}\right),~~ &\displaystyle 0<\theta<\theta_0 \\
	\end{array}\right.%
\end{align}
which is not continuous at $\theta_0$, we could measure the edge corrections to the torque, in the perfectly conducting limit, as
\begin{align}
	\left|\frac{\delta M}{M}\right|\approx0.066d_{3}\frac{P'(\theta)}{S'(\theta)}.
	\label{RC}
\end{align}
We see that the correction factor is a constant $0.264d_{3}/L$ for $\theta_0<\theta\leqslant\frac{\pi}{2}$. For the torque calculated in Fig.~\ref{TorqueAu2p}, the factor in this range is $0.00264\%$, and in the range of $0<\theta<\theta_0$, the largest relative correction is less than $0.0014\%$. This shows that the edge contribution to the torque is negligible. Though finite temperature will change the correction factor numerically, the conclusion remains true [At finite temperature, one could estimate the relative corrections from the Dirichlet boundary contribution~\cite{Weber} as an order-of-magnitude estimation~\cite{Gies}, which in the high temperature limit would result in a correction factor that is roughly ten times bigger than the one in Eq.~(\ref{RC})]. Notice that, strictly speaking, the above estimation may not be applied directly to the small-angle limit ($\theta\rightarrow0$), since in this circumstance our edge energy description in Eq.~(\ref{edgeE}) should be gradually changed to the energy form of the aligned edge. However, the aligned edge contribution is expected to be further reduced or at least not obviously strengthened~\cite{Gies}.

Qualitatively, from the viewpoint of van der Waals interactions, the edge contributions may also be ignored for a dielectric torque with ordinary settings, due to the simple fact that the rapid long-range decays of the van der Waals interactions between atoms or molecules in two dielectric plates should be the same as in conducting plates~\cite{Zhang}.

\section{Concluding remarks}
The magnetic response is a basic electromagnetic property of matter, and its nontrivial role in the Casimir effect is now widely appreciated. To include the influence of magnetic response, we extended the tangential Casimir force to magnetic media, which, in a general sense, provides all the necessary descriptions of this effect for isotropic media. Future experimental investigation of the force will require reliable theoretical predictions for guiding and comparison, and the temperature would be the only important residual ambient factor in such experiments. Thus, we considered the tangential forces at room temperature for gold, aluminum, and nickel plates, using their temperature-dependent optical data, by the Drude and the plasma treatments, respectively, and the needed information for measurements performed at room temperature was provided. Besides the tangential force, we investigated a related effect that the torque between isotropic plates in the parallel configuration, which offers a simple way to induce torques for uncharged surfaces.

\end{document}